\documentclass[11pt]{article}
\usepackage{scicite}
\usepackage{times}
\usepackage{amsmath}
\usepackage{physics}
\usepackage{pdfpages}
\usepackage{graphicx}
\usepackage[labelfont=bf]{caption}
\newenvironment{sciabstract}{%
\begin{quote} \bf}
{\end{quote}}
\usepackage[]{xcolor}
\usepackage{soul}
\usepackage{cancel}

\title{Strong coupling to circularly polarized photons: towards field-induced enantioselectivity}
\author
{Rosario R. Riso,$^{1}$ Enrico Ronca $^{2^\ast}$ and Henrik Koch$^{1,3^\ast}$\\
\normalsize{$^{1}$Department of Chemistry, Norwegian University of Science and Technology,}\\
\normalsize{7491 Trondheim, Norway}\\
\normalsize{$^{2}$Department of Chemistry, Biology and Biotechnology, University of Perugia,}\\
\normalsize{Via Elce di Sotto, 8, 06123, Perugia, Italy}\\
\normalsize{$^{3}$Scuola Normale Superiore, Piazza dei Cavalieri 7, 56126 Pisa, Italy}\\
\\
\normalsize{$^\ast$ E-mail:   enrico.ronca@unipg.it and henrik.koch@sns.it.}
}
\date{}
\begin{document}
\baselineskip24pt
\maketitle
\begin{sciabstract}
The development of new methodologies for the selective synthesis of individual enantiomers is still one of the major challenges in synthetic chemistry. Many biomolecules, and therefore many pharmaceutical compounds, are indeed chiral. While the use of chiral reactants or catalysts has led to significant progress in the field of asymmetric synthesis, a systematic approach applicable to general reactions has still not been proposed. In this work, we show that strong coupling to circularly polarized fields represents a promising alternative for reaching highly selective asymmetric synthesis in a non-invasive fashion. We demonstrate that the field induces stereoselectivity in the early stages of the chemical reaction, by selecting an energetically preferred direction of approach for the reagents.  
\end{sciabstract}
\section*{Introduction}
Asymmetric synthesis is a formidable challenge in chemistry, as demonstrated by the 2021 Nobel prize in chemistry to List and MacMillan \cite{wilson2005enantioselective,dehovitz2020static,notz2000catalytic,list2002proline,jen2000new}. Most biologically relevant molecules belong to a special class of systems that cannot be superimposed with their own mirror image. These molecules, known as chiral molecules, exist in two possible configurations called enantiomers, which have identical energy levels and share most physical properties. As a result, achieving a selective synthesis of only one of the two enantiomers is a complicated task and reactions often produce a mixture of enantiomers instead of an enantiomerically pure product. This is a major problem in the pharmaceutical industry since most often only one enantiomer has the desired healing behavior, while the other may be ineffective or even harmful\cite{zhao2020stereospecific}. Living organisms have developed remarkable techniques to produce only one of the two configurations through the use of enzymes and catalysts \cite{sims2022enantiomeric,shin1999asymmetric}. Therefore, the most successful research lines in asymmetric synthesis have so far mostly focused in this direction\cite{list2002proline,deepa2021recent}. However, catalysts are often reaction specific and a systematic approach is yet to be developed\cite{jen2000new}. Strong coupling between molecules and vacuum photons fields has recently emerged as a promising technique to engineer molecular properties in a non-invasive way\cite{xiang2020intermolecular,mueller2020deep,bloch2022strongly,scholes2020polaritons,ashida2021cavity}. When light and matter interact strongly, for example inside optical cavities, they form new hybrid electron-photon states with unique properties, the polaritons \cite{hubener2021engineering,li2022molecular,weight2023investigating}. Intriguing applications of polaritonic chemistry include modifying molecular absorption and fluorescence\cite{herrera2017absorption,wang2020coherent,li2021cavity}, altering photochemical processes \cite{fregoni2018manipulating} and even speeding up or slowing down chemical reactions\cite{schafer2022shining,lather2019cavity,thomas2019tilting,ahn2023modification}. In a previous work, we demonstrated that circularly polarized fields can induce energy differences between enantiomers, even in the ground state. This result stems from the fact that the two enantiomers exhibit differential absorption with respect to left‐ and right-circularly polarised light (LHCP and RHCP, respectively), a phenomenon known as circular dichroism. Notably, these energy differences can be exploited to create enantioselective signatures in the rotational spectra of the two mirror images\cite{riso2023strong}.\\
In this work, we investigate whether the enantiomeric energy differences in chiral cavities can induce enantioselectivity in reactions that are normally non-selective. Indeed, while the effect on the formed enantiomer is intriguing in itself, an even more captivating prospect is whether the field can bias non-chiral reagents towards favoring the formation of one configuration over the other (see Fig. \ref{fig:Figure_1}a).  
In particular, we suggest that appropriate engineering of the field can favor a desired approach direction of the reagents, leading to the preferential formation of only one of the two possible products \cite{riso2022characteristic,haugland2021intermolecular,saez2018organic}. Furthermore, since strong coupling to quantized fields induces long-range correlation between molecules, we expect that the field footprint becomes increasingly relevant at large distances where the electronic effects become smaller and smaller. Our studies suggest that harnessing the power of quantum fields may hold the key to achieving high levels of selectivity in asymmetric synthesis. Specifically, we confirm that field discrimination has an increasingly larger role at long distances while at short ranges the Coulomb interaction dominates.  
\section*{Results and discussion}
\begin{figure}
    \centering
    \includegraphics[width=\textwidth]{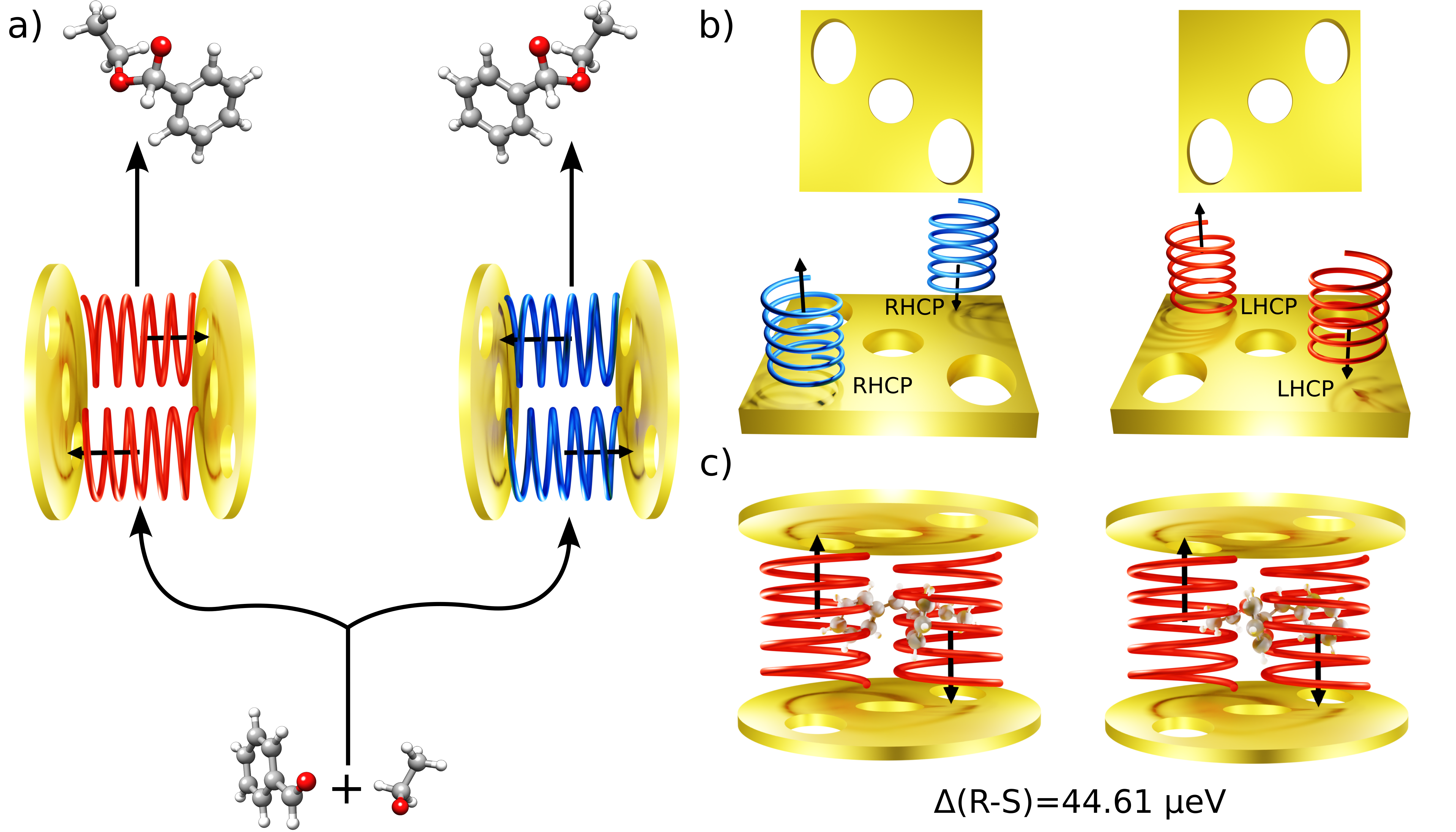}
    \caption{\textbf{Chiral cavities and their discriminating power.} \textbf{a}, Pictorial representation of a benzaldehyde reacting with ethanol inside an LHCP (red on the left) and RHCP (blue on the right) chiral cavity. Due to the field effects the reaction product has a different chirality. \textbf{b}, Example of mirrors reflecting only RHCP and LHCP light without changing the field circular polarization, as reported by Baranov and co-workers \cite{taradin2021chiral,voronin2022single}. \textbf{c}, Energy difference between two enantiomers of 1-ethoxy-1,2-diphenylethanol in a LHCP cavity. }
    \label{fig:Figure_1}
\end{figure}
Strong coupling between molecules and circularly polarized fields can be achieved using chiral cavities, devices that confine only one circular polarization of the field within a reduced volume, known as the quantization volume\cite{schafer2023chiral,baranov2023toward,sun2022polariton,mauro2023chiral}. Experimental realization of such devices has been reported by Gautier \textit{et al.} \cite{gautier2022planar} as well as by Baranov and co-workers \cite{taradin2021chiral,voronin2022single}, see Fig.\ref{fig:Figure_1}b for a pictorial representation. 
Exploiting the unique properties of the circularly polarized fields, chiral cavities effectively break the energy degeneracy between the two mirror images of a chiral molecule. To illustrate this phenomenon, we report in Fig. \ref{fig:Figure_1}c the field-induced energy difference between the two enantiomers of a chiral molecule, 1-ethoxy-1,2-diphenylethanol, placed inside a LHCP chiral cavity. The field frequency is set to $\omega$\;=\;2.7 eV and the light-matter coupling is $\lambda$\;=\;0.05 atomic units (a.u.). Detailed information regarding these parameters and their physical meaning can be found in the Methods. In line with Ref.\cite{riso2023strong}, we refer to the enantiomeric energy differences induced by the cavity as the field discriminating power. 
For the 1-ethoxy-1,2-diphenylethanol molecule in Fig.\ref{fig:Figure_1}c, the ground state field discriminating power is on the order of 40 $\mu$eV, consistent with the findings reported previously in the literature \cite{riso2023strong,ke2022can}. Even though this energy range can be experimentally detected using modern instruments (e.g. nuclear magnetic resonance operates in a similar energy range), it is important to note that the chiral effects are several orders of magnitude smaller than both room temperature thermal energy and molecular binding energies, respectively on the order of 25 meV and hundreds meV. Although the field-induced discrimination increases for larger chiral systems and stronger light-matter coupling strengths, the magnitude of the discriminating power remains within the $\mu$eV range per molecule for real systems. The effects can however be enhanced up to the chemically relevant range (meV) when increasing the number of chiral molecules inside the cavity. \\
\subsection*{Benzaldehyde-ethanol reaction}
\begin{figure}
    \centering
    \includegraphics[width=\textwidth]{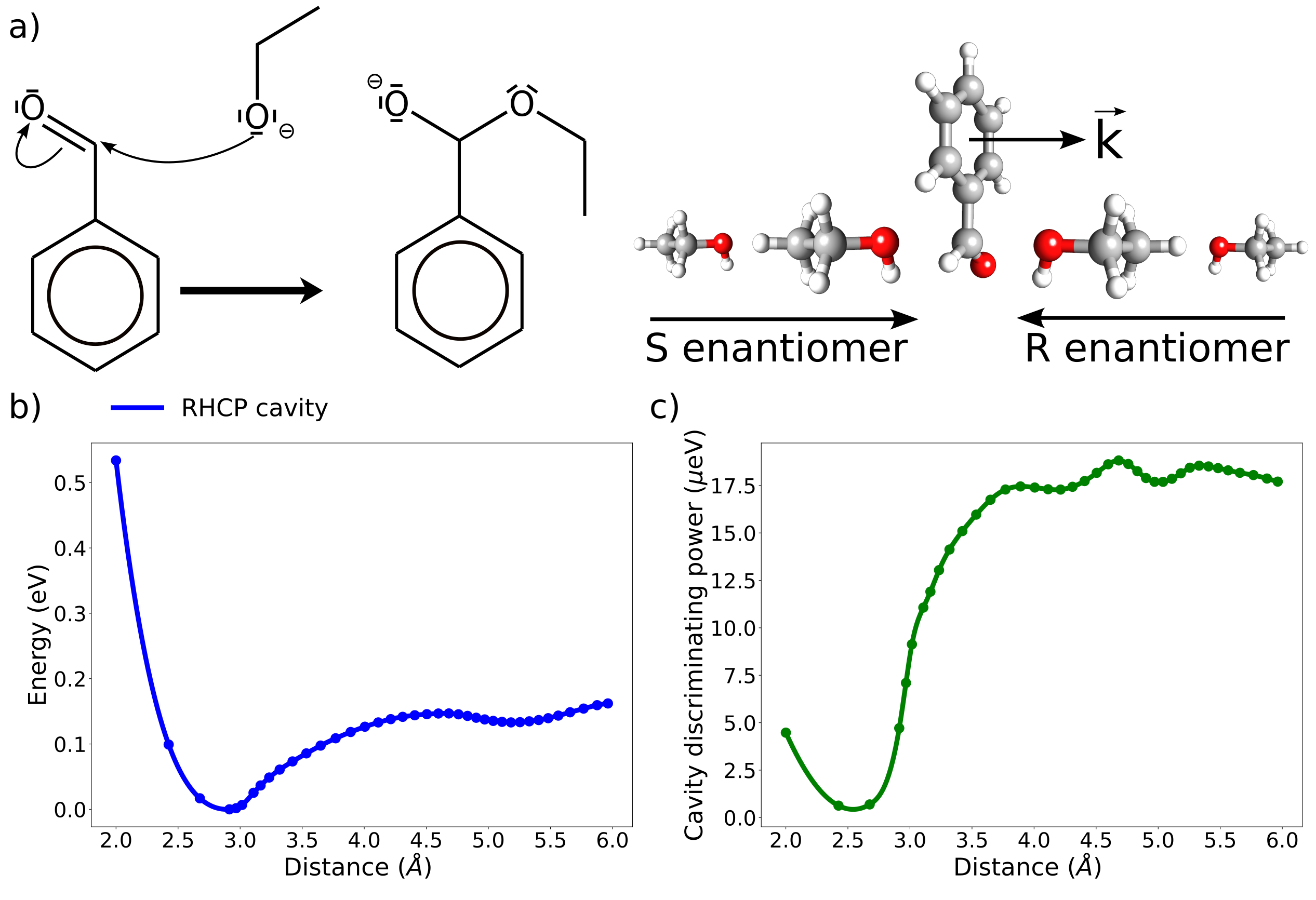}
    \caption{\textbf{Field effects on the short-range benzaldehyde-ethanol reaction.} \textbf{a}, Reaction mechanism and stereoselectivity for the benzaldehyde-ethanol reaction. \textbf{b}, Potential energy surface for the reaction, the ethanol is approaching parallel to the cavity wave vector $\mathbf{k}$. \textbf{c}, Cavity discriminating power, computed by subtracting the potential energy surfaces for right and left approaches, see panel \textbf{a}. We notice that the sign of the field discrimination power remains constant along the full pathway.}
    \label{fig:Benzaldeide}
\end{figure}
\begin{figure}
    \centering
    \includegraphics[width=\textwidth]{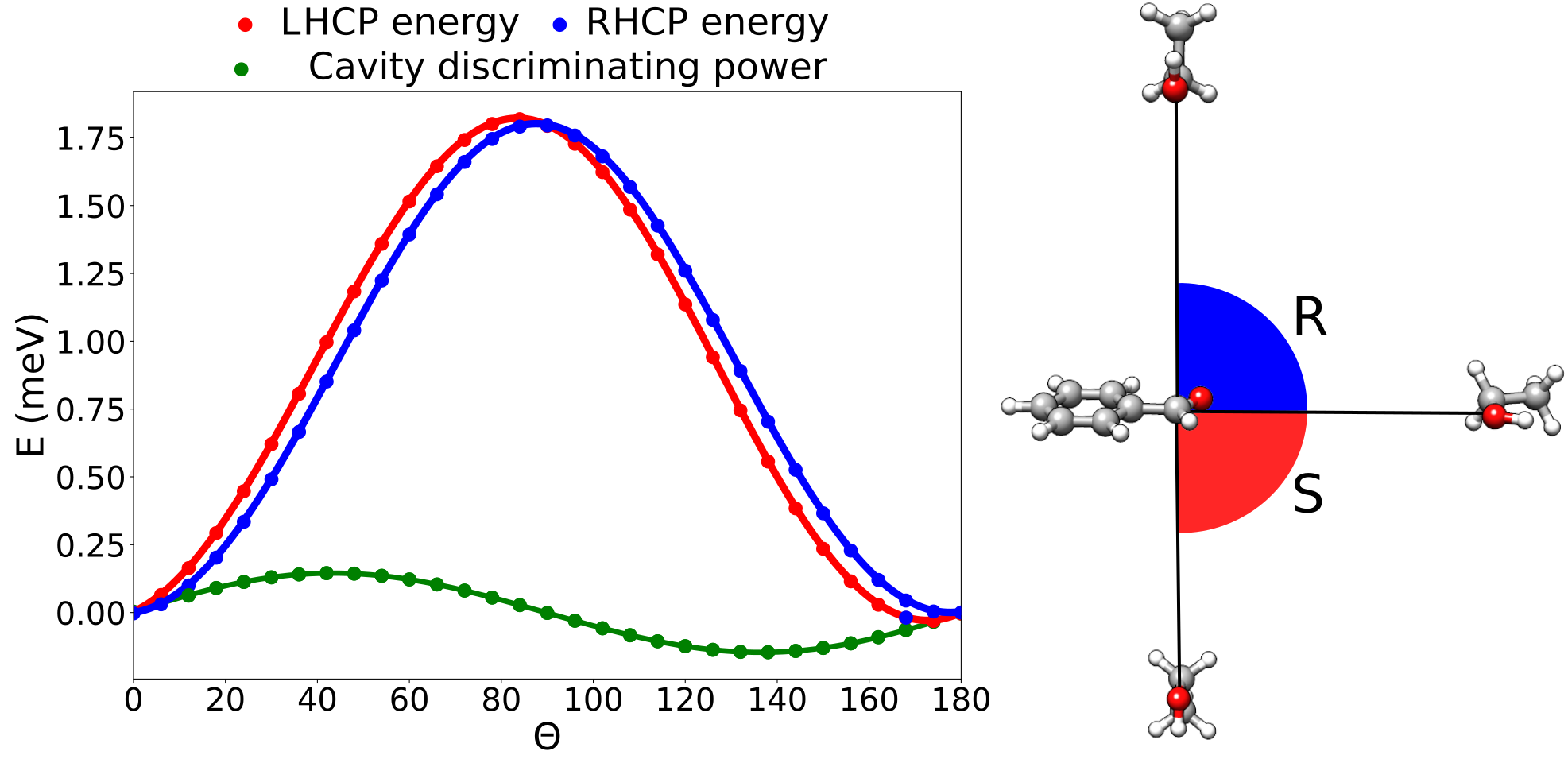}
    \caption{\textbf{Angular dispersion of the energy for the hemiacetal reaction with the reagents placed 200 \AA\; apart.} The sign change in the chiral discriminating power shows the inversion of the preferred chirality between LHCP and RHCP cavities.}
    \label{fig:Rotation_and_long_range}
\end{figure}

In Refs.\cite{riso2023strong,castagnola2023polaritonic}, we show that since the cavity field introduces a spacial anisotropy, the molecular energy is highly dependent on the system orientation. As a consequence, during a reactive event every reaction pathway is either stabilized or destabilized by the field\cite{pavovsevic2023computational,pavosevic2023cavity}. This observation suggests that strong coupling to circularly polarized fields has the potential to influence the stereoselectivity  of chemical reactions by favoring specific approach directions. To investigate this idea, we focus on the field effects on the non-enantioselective reaction between benzaldehyde and ethanol leading to the formation of a hemiacetal (see Fig.\ref{fig:Benzaldeide}a). The approach direction of the reagent plays a crucial role in determining the chirality of the final product. Specifically, the orientation of the alcohol group relative to the aromatic ring is the key variable. A crossing of this plane results in a chirality inversion of the product. 
In Fig.\ref{fig:Benzaldeide} we illustrate the cavity effect on two different approach directions of the alcohol group: from left and right with respect to the aromatic ring plane. In particular, the benzaldehyde and the ethanol approach each other parallel with respect to the field wave vector $\mathbf{k}$. %
The chiral cavity utilized in these calculations is a RHCP cavity with an optical frequency of $\omega = 2.7$ eV and a coupling strength of $\lambda = 0.05$ a.u. .
In Fig.\ref{fig:Benzaldeide}b, we show the potential energy surfaces (PES) of the reaction while in Fig.\ref{fig:Benzaldeide}c we report the energy difference between the two approach directions, (left minus right), as a function of the distance. We point out that in the non-cavity case, the energy difference in Fig.\ref{fig:Benzaldeide}c would be zero. The chiral field can instead discriminate between the two approach directions leading to a net difference in the PES profile. These field-induced energy differences between reactive pathways leading to two different enantiomers is exactly the effect this paper focuses on. The sign of the field stabilization is constant along the reactive path, indicating that the cavity consistently favors the right approach of the alcohol in Fig.\ref{fig:Benzaldeide}a, favouring a R enantiomer product. However, the electronic effects overwhelmingly dominate compared to the contributions from the cavity, which remain in the $\mu$eV range, and only a negligible fraction of the reactive pathways will be redirected due to the effects discussed in Fig.\ref{fig:Benzaldeide}.
Yet we notice that the cavity stabilization of the approach creating a R enantiomer increases as the distance grows. This is a crucial observation because while the electronic effects diminish quite rapidly with distance, strong coupling to quantized fields introduces long-range correlation between molecules. In an ideal gas phase experiment, where the reagents are approaching from a large distance, the discriminating power of chiral fields might therefore become the dominating effect.
To exemplify the persistence of the cavity discriminating power for large separations, we illustrate the field effects for the benzaldehyde-ethanol reaction when the reagents are 200 \AA\; apart. Specifically, Fig.\ref{fig:Rotation_and_long_range} displays the angular dispersion of the energy for a RHCP and a LHCP cavity, with the field-induced effects computed as the difference between the red and the blue curves.
The discriminating power of the cavity (obtained subtracting the LHCP results with the RHCP results) exhibits a prominet sign change around the $90^{\circ}$ mark, i.e. where the system's chirality is inverted. Specifically, RHCP fields stabilize the R enantiomer ($\theta$ less than $90^{\circ}$) more than the S enantiomer ($\theta$ larger than $90^{\circ}$), while the opposite behavior is observed in the LHCP cavity. This result validates the intuitive picture that when the system inverts its chirality, the field stabilization also changes sign.
Although the cavity-induced discrimination remains in the $\mu$eV range, it is crucial to consider that in Fig.\ref{fig:Rotation_and_long_range} the Coulomb interactions between the reagents are negligible and the field effects can accumulate over the whole approach path. The shape of the angular dispersion is predominantly determined by dipole effects, which are, in agreement with the theoretical description, larger than the enantiomeric discrimination. 
However, at these large distances, the enantioselective field effects are comparable to the dipole effects and the dispersion curves for the rotation in the RHCP and LHCP cavities are therefore qualitatively different.  
\begin{figure}
    \centering
    \includegraphics[width=\textwidth]{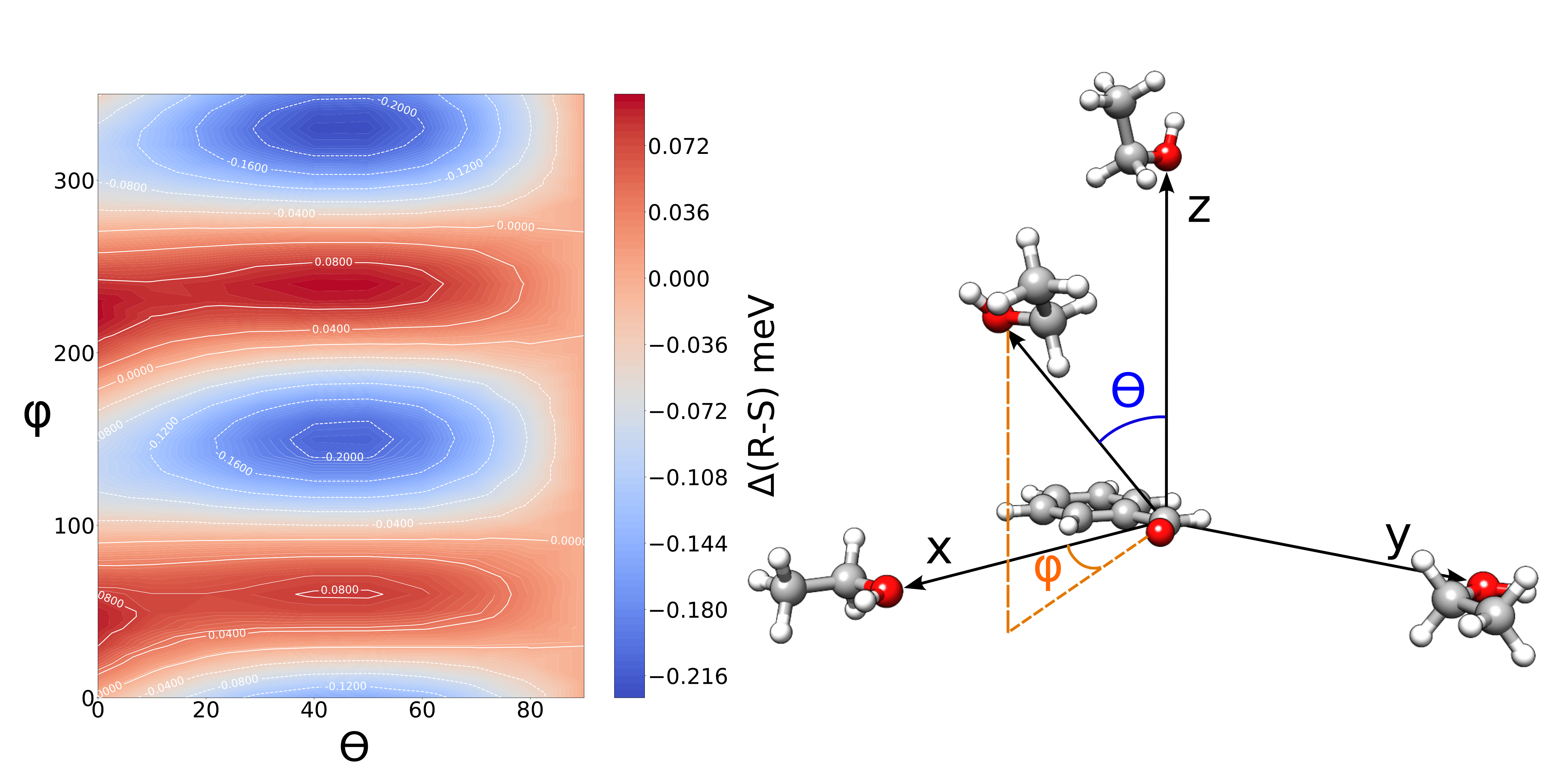}
    \caption{\textbf{Field-induced stabilization of the R (blue) or S (red) enantiomers as a function of the relative orientation between the two reagents.}}
    \label{fig:200AA}
\end{figure}
In the long distance range, it is critical to account for orientational effects as the molecules are free to reorient with respect to each other. To determine whether the orientational effects will cancel the long-range contributions of the cavity discriminating power, we present in Fig.\ref{fig:200AA} the field-induced energy differences between the top and bottom reactions for all possible approach directions of the alcohol in a RHCP cavity, as obtained by varying the spherical angles $\theta$ and $\phi$. 
The plot in Fig.\ref{fig:200AA} clearly presents peaks and valleys, with the deepest minimum (in blue) corresponding to the path where the R enantiomer formation is maximally favoured, and the highest maximum (in red) indicating the approach direction that most significantly favours the S enantiomer. Furthermore, we notice that the minimum is more pronounced than the maximum. When a reaction is carried out inside a chiral cavity, like in Fig.\ref{fig:200AA}, every configuration of the potential energy surface is populated following the Boltzmann distribution 
\begin{equation}
P(\theta,\phi)=\frac{\textrm{exp}\left(-E(\theta,\phi)/k_{b}T\right)}{\sum_{\theta,\phi}\textrm{exp}\left(-E(\theta,\phi)/k_{b}T\right)},  
\end{equation}
where $P(\theta,\phi)$ is the probability of finding the molecule in a certain configuration, $T$ is the reaction temperature and $k_{b}$ is the Boltzmann constant. 
\begin{figure}
    \centering
    \includegraphics[width=\textwidth]{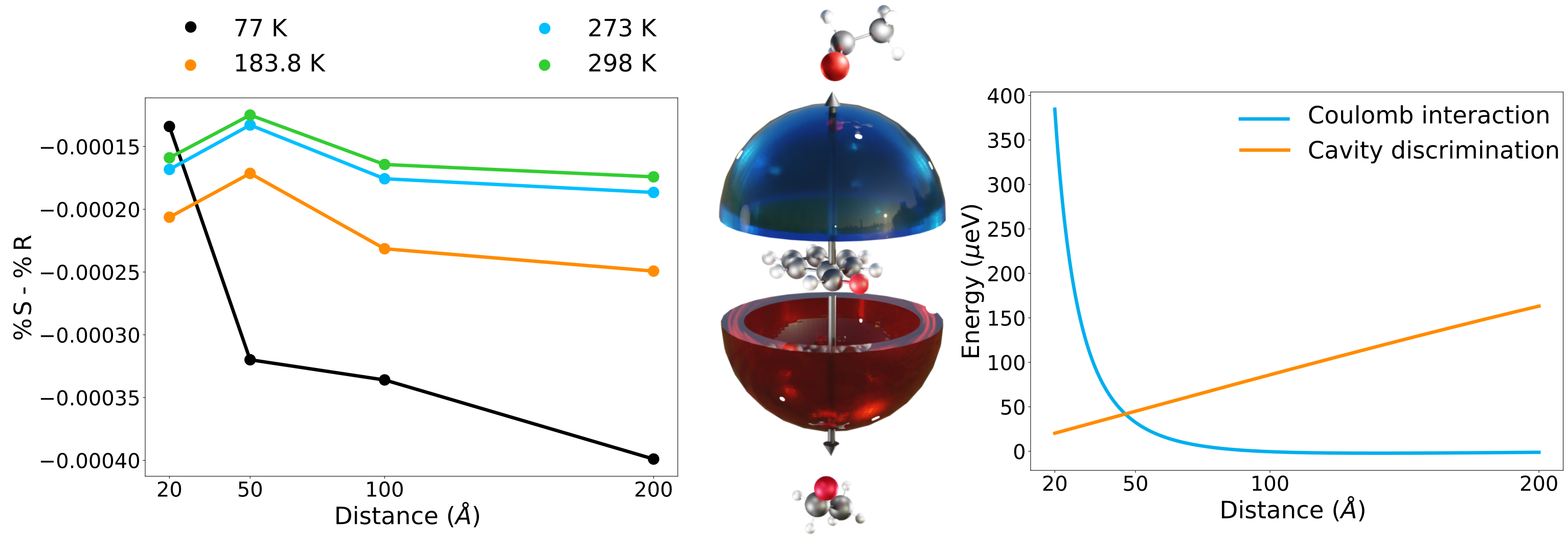}
    \caption{\textbf{Field-induced enantioselectivity as a function of the distance and of the reaction temperature.} We notice that the enantioselectivity always displays the same sign, becoming more pronounced for larger distances. On the right we show how the Coulomb contributions decay much faster than the field effects.}
    \label{fig:Boltzmann}
\end{figure} 
In Fig.\ref{fig:Boltzmann} we therefore compare the probability of an approach from above (blue hemisphere) with an approach from below (red hemisphere) for the benzaldehyde-ethanol reaction at different distances. To check the robustness of the results over the reaction parameters, we consider the reactive approach at different temperatures: 298 K (room temperature), 273 K ($0\;^{\circ}$C), 183 K (lowest temperature ever measured on earth) and 77 K (nitrogen condensation).
The results in Fig.\ref{fig:Boltzmann} demonstrate that the field-induced effects persist even after rotational averaging. The same enantiomer is favored at all distances and for all temperatures confirming that the cavity discriminating power is robust. Furthermore, the field discrimination is more pronounced at larger distances. This is a crucial observation because, as the Coulomb interaction becomes negligible the chirality effects dominate the interaction between the fragments. Specifically, on the right side of Fig.\ref{fig:Boltzmann} we plot the magnitude of the Coulomb and chiral interactions for the approach at $\theta=40^{\circ}$ and $\phi=20^{\circ}$, see Methods section. We observe that the field-induced discrimination is significantly larger than the Coulomb contribution before 50 \AA. As expected, the cavity effect on the stereoselectivity increases as the temperature decreases because the lower energy approach direction is more populated at lower temperatures. The field-induced enantioselectivity reported in Fig.\ref{fig:Boltzmann} may be small, with only a few fractions of a percentage gained even at 77 K. However, it is important to note that the effect is not zero even at very large distances, meaning that accumulation of the chirality effects needs to be taken into account. We firmly believe that with careful engineering of the field and reaction conditions, a significant impact on the chirality of the final product can be achieved.
\subsection*{Origin of the field-induced discrimination}
In Ref.\cite{riso2023strong}, we demonstrate that the field-induced discrimination of a chiral cavity is directly linked to the optical activity of the molecule. This observation substantiates the claim that field-induced energy discrimination arises due to the molecular circular dichroism. This is a remarkable result since it connects to the idea of homochirality, where the field promotes the enantiomer with the same chirality \cite{baranov2023toward}. A more detailed investigation of the homochirality hypothesis in chiral cavities will be the subject of a future work as our findings indicate that correlation between optical rotatory power and field stabilization is not straightforward (only the positive pole of the optical activity determines the effect). Even so, given a certain reaction, it is possible to choose the field polarization based on the desired enantiomer we wish to produce. 
The field discriminating effects can also be attributed to a different mechanism, well established in the chemistry community. When dealing with racemic mixtures containing both enantiomers, one effective approach for their separation involves the reaction of the mixture with an enantiomerically pure system. Despite still being isomers, the two reaction products are indeed no longer mirror images of one another; they are instead diastereoisomers. Diastereoisomers have different physical and chemical properties and therefore techniques such as distillation, crystallization, chromatography, or extraction can be employed to separate them\cite{vu2022enhanced}. Nowadays this procedure is not used as much because the reactive steps significantly reduce the yield of the separation process. The field discriminating power in chiral cavities parallels the diastereomeric mechanism described above. Specifically, while the two enantiomers of a chiral molecule have the same energy outside chiral cavities, upon interaction with a chirally pure entity, the circularly polarized field in our framework, two different diastereoisomers are formed. Inside a chiral cavity, no symmetry operation can interconvert one enantiomer into the other\cite{riso2023strong}. While drawing an analogy between light-matter interaction and chemical bonding might appear bold, in the strong coupling regime the field effects often mirror molecular interactions, see the solvent caging effect reported by Li \textit{et al} \cite{li2021theory,li2021cavity}. \\
\section*{Discussion}
Regardless of the conceptual framework chosen to rationalize the discriminating power of the field, our work demonstrates that strong coupling between molecules and tightly confined circularly polarized fields has the potential to affect reactivity, eventually leading to enantioselective synthesis. Despite being small compared to the electronic effects when the reagents are close, the cavity discriminating power becomes increasingly relevant at large separations. The field does indeed create long-range correlation between the fragments that can reorient at long distances. Overall,
this research opens the way to new and unexplored paths toward asymmetrical synthesis, with potential implications across multiple areas, from drug design to material science. Considering the recent fast advances made from the experimental side in the fabrication of chiral cavities \cite{gautier2022planar,taradin2021chiral,voronin2022single,plum2015chiral}, we believe that the proposed methodology will, in a near future, find an experimental validation. Future efforts will be devoted to studying the field-induced discrimination for excited states with a particular focus on collective effects\cite{baranov2023toward,schafer2023chiral,sidler2022perspective}. Additionally, the study of strong coupling to circularly polarized fields in the vibrational energy\cite{mandal2022theory} range will be the subject of further investigations. 
\section{Acknowledgements}
R. R. R and H. K. acknowledge funding from the
Research Council of Norway through FRINATEK Project
No. 275506. This work has received funding from the European Research Council (ERC) under
the European Union’s Horizon 2020 Research and Innovation Programme (Grant Agreement No. 101020016). E. R. acknowledges funding from the European Research Council (ERC) under the European Union’s Horizon Europe Research and Innovation Programme (Grant No. ERC-StG-2021-101040197—QED-Spin).
\section{Author contributions}
R.R.R., E.R. and H.K. developed the theory. E.R. conceived the project. R.R.R.
implemented the computational methodology in the eT program. All the authors proposed the investigated applications. R.R.R. obtained the data. R.R.R. prepared a first draft of the paper. R.R.R. and E.R. worked at the graphical part of the manuscript. H.K. and E.R. oversaw the project. All the authors discussed the results and edited the manuscript.
\section{Competing interests}
The authors declare no competing interests.
\section{Data availability statement}
The geometries used for the reported calculations can be found on the Zenodo open repository \cite{rosario_roberto_riso_2023_8208785}. The eT program\cite{folkestad20201} is open source and instructions on how to install can be found on the Zenodo link. Additional data on a SN1 reaction, on the cavity effects on Van Der Waals complexes and on the theoretical methodology used to compute the results of this paper can be found in the Supplementary Materials. 

\clearpage


\section*{Supplementary material}

# Supplementary materials to:
# Strong coupling to circularly polarized photons: towards field-induced enantioselectivity

Rosario R. Riso,[1] Enrico Ronca[2*] and Henrik Koch[1,3*]

[1]Department of Chemistry, Norwegian University of Science and Technology,
7491 Trondheim, Norway

[2]Department of Chemistry, Biology and Biotechnology, University of Perugia,
Via Elce di Sotto, 8, 06123, Perugia, Italy

[3]Scuola Normale Superiore, Piazza dei Cavalieri 7, 56126 Pisa, Italy

* E-mail: enrico.ronca@unipg.it and henrik.koch@sns.it.

## 1   Materials and Methods

The interaction between the electrons and the field is modeled using the Born-Oppenheimer minimal coupling Hamiltonian

$$H = \frac{1}{2} \sum_i \left(\mathbf{p}_i - \mathbf{A}(\mathbf{r}_i)\right)^2 + \sum_{i>j} \frac{1}{|r_i - r_j|} + \sum_{I>J} \frac{Z_I Z_J}{|R_I - R_J|} - \sum_{i,I} \frac{Z_I}{|R_I - r_i|} \\ + \frac{1}{2} \int \left(\mathbf{E}^2(\mathbf{r}) + c^2 \mathbf{B}^2(\mathbf{r})\right) d^3 r, \quad (1)$$

where $i$ and $j$ label electrons while $I$ and $J$ label nuclei with charges $Z_I$ and $Z_J$. The vector potential, the electric and the magnetic fields are denoted by $\mathbf{A}(\mathbf{r})$, $\mathbf{E}(\mathbf{r})$ and $\mathbf{B}(\mathbf{r})$, respectively. In the strong light-matter coupling regime, the field is a critical component of the system and therefore the quantum nature of the photons needs to be taken into account. This is achieved by adopting a quantum electrodynamical (QED) description. Inside a chiral cavity, the field is circularly polarized, i.e. the second quantized vector potential equals (*1–3*)

$$\mathbf{A}(\mathbf{r}) = \sum_{\mathbf{k}} \frac{\lambda}{\sqrt{2\omega_k}} \left(\boldsymbol{\epsilon}_{\mathbf{k}\pm} e^{i\mathbf{k}\mathbf{r}} \left(\alpha_{\mathbf{k}} + \beta^\dagger_{\mathbf{k}}\right) + \boldsymbol{\epsilon}^*_{\mathbf{k}\pm} e^{-i\mathbf{k}\mathbf{r}} \left(\alpha^\dagger_{\mathbf{k}} + \beta_{\mathbf{k}}\right)\right), \quad (2)$$

1

where $\alpha/\beta$ and $\alpha^\dagger/\beta^\dagger$ respectively create and annihilate a photon with wave vector **k** and polarization $\epsilon_{\mathbf{k}\pm}$. The $\pm$ subscript in Eq.2 refers to the circular polarization of the field, + for LHCP light and - for RHCP. The photon frequency $\omega$ is related to the wave vector via the dispersion relation $\omega = \frac{k}{c}$ while the $\lambda$ parameter quantifies the strength of the light-matter interaction. Specifically, when $\lambda = 0$ a.u., the field and the molecule are completely decoupled while the field-induced effects become increasingly relevant for larger $\lambda$. For this reason, recent efforts are devoted towards the fabrication of cavities with higher coupling strengths. Since $\lambda = \sqrt{\frac{\hbar}{\epsilon_0 V}}$, stronger couplings are usually obtained by extreme confinement of the quantization volume, for example in picocavities of nanoplasmonic devices (*4, 5*). Experiments keep moving the limit of ultrastrong coupling with the recent realizations of subnanometric cavities, i.e. $\lambda \approx 0.05$ a.u. . We point out that two photons are needed in order to consistently describe the vector potential in Eq.2. The wave function is modeled using the minimal coupling quantum electrodynamics coupled cluster methodology (MC-QED-CCSD-SD) (*3, 6*) implemented in a development version of the eT program (*7*)

$$|\psi\rangle = \exp\left[T_1 + T_2 + (S_{2\alpha} + S_{1\alpha} + \gamma_\alpha)\alpha^\dagger + (S_{2\beta} + S_{1\beta} + \gamma_\beta)\beta^\dagger\right] |HF\rangle \otimes |0,0\rangle, \qquad (3)$$

where, in addition to the usual electronic excitation operators $T_1$ and, $T_2$, electron-photon and photon excitation operators are included in the cluster operator

$$\begin{aligned}
T_1 &= \sum_{ai} t_i^a E_{ai} & T_2 &= \frac{1}{2}\sum_{abij} t_{ij}^{ab} E_{ai} E_{bj} \\
S_{1\alpha} &= \sum_{ai} s_{i\alpha}^a E_{ai} & S_{1\beta} &= \sum_{ai} s_{i\beta}^a E_{ai} \\
S_{2\alpha} &= \frac{1}{2}\sum_{aibj} s_{ij\alpha}^{ab} E_{ai} E_{bj} & S_{2\beta} &= \frac{1}{2}\sum_{aibj} s_{ij\beta}^{ab} E_{ai} E_{bj}.
\end{aligned} \qquad (4)$$

In Eq.4, HF denotes the Hartree-Fock Slater determinant and 0,0 is the photonic vacuum on both $\alpha$ and $\beta$ photons. The electronic operators $E_{ai}, E_{bj}$ excite an electron from the occupied orbitals $i, j$ into the virtual orbitals $a$ and $b$. A detailed discussion of the theoretical description of chiral cavities can be found in Ref. (*3*). All the calculations have been performed using a cc-pvdz basis set (*8, 9*). The geometries

used for the reported calculations have been optimized using the Orca package at the B3LYP/def2-SVP level (*10*). Moreover, the nudged elastic band approach (*11*) has been used to compute the reaction path for the benzaldehyde reaction in Fig.2. In Fig.5, the Coulomb contribution has been computed at the CCSD level subtracting the energy of the individual fragments from the energy of the two fragments at every distance.

## 2   Van der Waals compounds

In most cases, enantiomers cannot be converted into each other without breaking chemical bonds. Therefore, the equilibrium composition of an enantiomeric mixture within a chiral cavity is not significantly modified due to the field-induced energy differences in the ground state. However, the cavity effect should be significantly more pronounced for chiral molecules that can undergo interconversion without the need for breaking bonds. As a case study, we consider van der Waals complexes, stable systems that are held together by intermolecular forces, i.e. dispersion (*12*). Specifically, the system we focus on is composed of two substituted acetylene compounds forming a cross in the most stable configuration. Rotation of either one of the two acetylene compounds along the intermolecular axis leads to enantiomeric interconversion, see Fig.1a. The energy profile of this rotation in vacuum (black) and in a LHCP cavity (blue) is shown in Fig.1b. In both cases a peak in the angular dispersion, representing the interconversion barrier, is observed at approximately $90°$. Notably, the barrier height more than doubles inside the chiral cavity increasing from 0.04 eV up to 0.1 eV. These cavity-induced effects are consistent with findings from previous literature studies (*13–15*). If the chiral field contributions are isolated (subtracting for a given angle $\theta$ the energy between the molecule and its mirror image), the dispersion curve in Fig.1c is obtained. The observed trend exhibits a striking similarity to the one already discussed in Fig.3. Specifically, we recognize the sign change in the field stabilization at the $90°$, i.e. where the system's chirality is inverted. The field stabilizes the R enantiomer ($\theta$ larger than $90°$) more than the S enantiomer ($\theta$ smaller than $90°$) in a RHCP cavity. If the number of strongly coupled complexes is increased, the chirality dis-

3

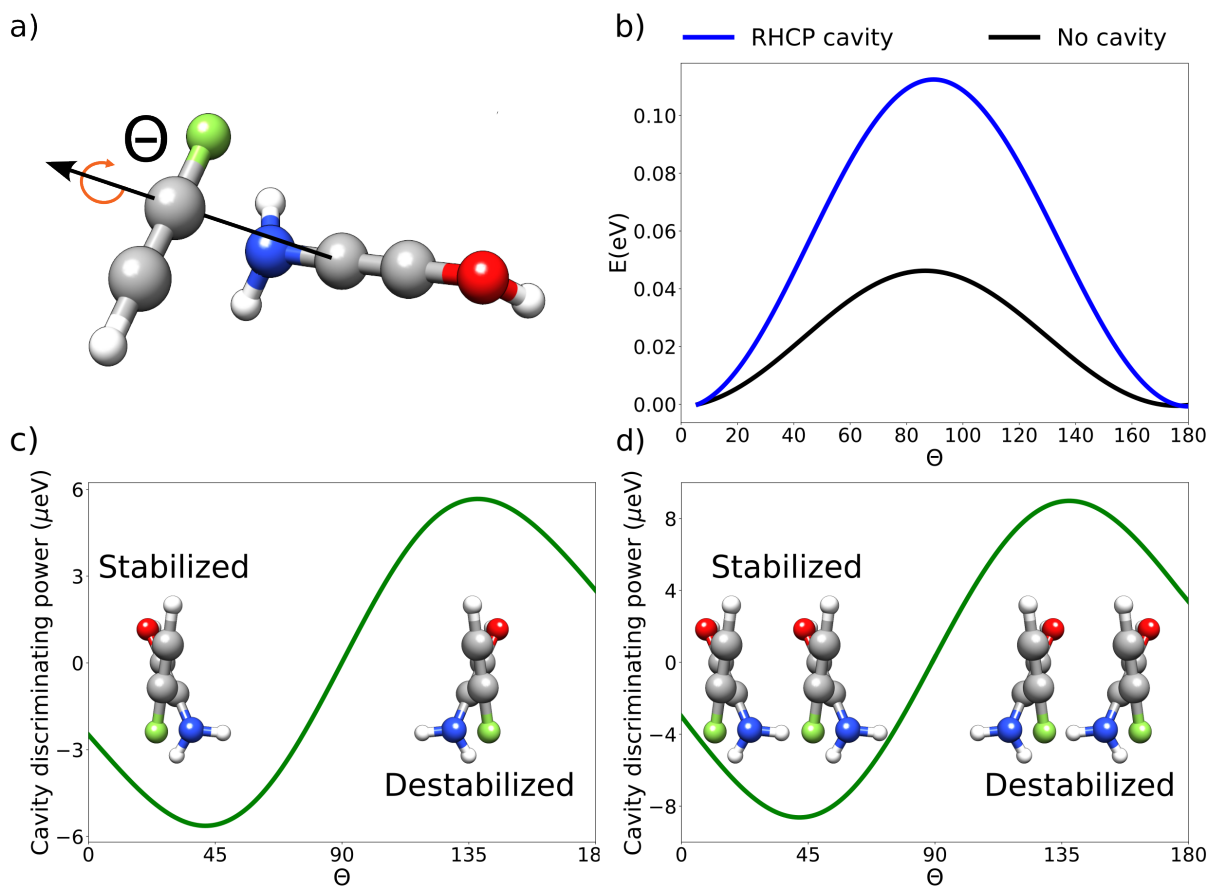

Figure 1: **Cavity induced discrimination on a van der Waals chiral compound. a-b** Potential energy surface for the enantiomeric interconversion inside and outside the cavity as obtained varying the angle $\theta$. Strong coupling to the field induces a significant change in the energy profile doubling the barrier height. **c-d** Field stabilization as a function of the interconversion reaction coordinate for one or two van der Waals compounds. We note that the cavity discrimination changes sign as the complex changes its chirality and that the effect increases when the number of strongly coupled molecules increases.

4

criminating effects also increases, as shown in Fig.1d and discussed in Ref. (*3*). The chiral field effects are still several orders of magnitude smaller than the energetic barrier, but their relevance in percentage is significantly larger than for normal reactive events. Moreover, the complex shown in Fig.1 is held together by dipole-dipole type interactions, i.e. the strongest intermolecular interactions that do not involve charges or hydrogen-bonds. In other van der Waals complexes held together only by dispersion forces, the field effect should be even more pronounced.

## SN1 reaction and role of the electronic wave function

In the main text, we discussed the field effects on the reactivity of the hemiacetal reaction. In this section, we study the SN1 reaction between a cation of N,N-dimethyl neopentyl amine and methanol in a RHCP cavity. The cavity parameters are $\omega$ = 2.7 eV and $\lambda$ = 0.05 a.u. . The reagents approach each other in the direction perpendicular to the field vector **k**. Similarly to the benzaldehyde case in the main text, we notice that the cavity discriminating power has the same sign along the reactive pathway, decreasing as the distance increases. Moreover, we observe that both the potential energy surface and the field-induced discrimination exhibit a discontinuity at around 4 Å. Discontinuities in the potential energy surfaces are a well-known problem in electronic structure methods related to an abrupt change in the system wave function with the creation or breaking of chemical bonds. These discontinuities can be eliminated using a complete active space approach (CAS) (*16, 17*). Future investigations will focus on the development of QED-CAS methodology. However, it is important to note that for the SN1 reaction in Fig.2, the discontinuities do not affect the qualitative conclusions drawn in this work. Moreover, since the reagents are closed shell, there is no multireference character in the wave function at long range. Furthermore, we highlight that discontinuities in the cavity discriminating power coincide with the discontinuities in the potential energy surface. The chirality effect of the field, therefore, is sentive to the system wave function and not only the approach geometry (which changes smoothly for the reaction in Fig.2).

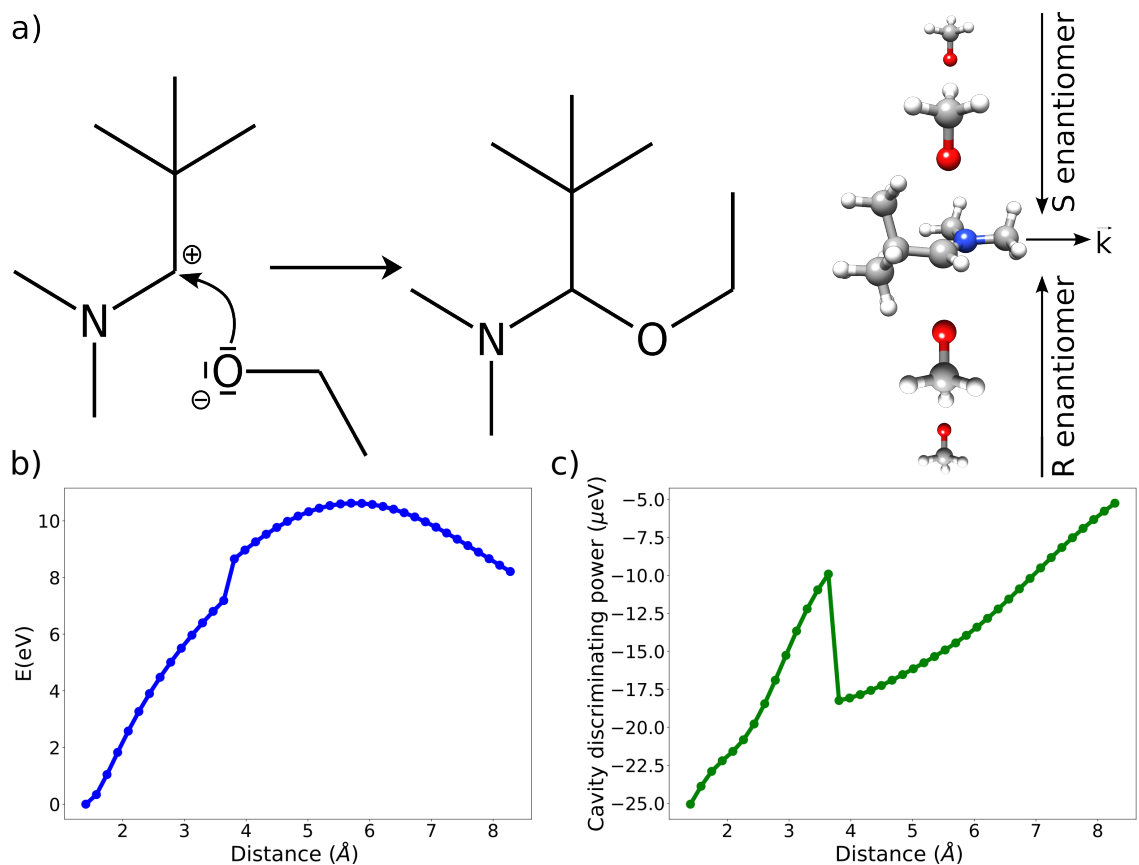

**Figure 2: Field effects on a short-range SN1 reaction. a**, Reaction mechanism and stereoselectivity for the N,N-dimethyl neopentyl amine and methanol reaction. **b**, Potential energy surface for the reaction, the methanol is approaching perpendicular to the cavity wave vector **k**. **c**, Cavity discriminating power, computed by subtracting the potential energy surfaces for top and bottom approaches, see panel **a**. We notice that the sign of the field discrimination power is the same along the full pathway.

6


\end{document}